\documentclass[preprint,number]{elsarticle}

\usepackage{graphicx}
\usepackage{amsmath}

\def\cm{\ref{eqn:dtheta}--\ref{eqn:dn}}
\def\citemn{\cite{2009arXiv0904.3253M,newman2009random}}

\begin{document}

\begin{frontmatter}
\author[ume]{Erik M Volz}
\ead{erikvolz@umich.edu}
\ead[url]{http://www.erikvolz.info}
\address[ume]{Department of Epidemiology, University of Michigan--Ann Arbor,
1415 Washington Heights, Ann Arbor, MI 48109-2029}

\title{Dynamics of infectious disease in clustered networks with arbitrary degree distributions }
\date{\today}
\begin{abstract}
We investigate the effects of heterogeneous and clustered contact patterns on the timescale and final size of infectious disease epidemics. 
The abundance of transitive relationships (the number of 3 cliques) in a network and the variance of the degree distribution are shown to have large effects on the number ultimately infected and how quickly the epidemic propagates. 
The network model is based on a simple 
 generalization of the configuration model, and
 epidemic dynamics are modeled with a low dimensional system of ordinary differential equations. 
Because of the simplicity of this model, we are able to explore a large parameter space and characterize dynamics over a wide range of network topologies.
We find that the interaction between clustering and the degree distribution is complex, and that clustering always slows down an epidemic, but that simultaneously increasing clustering and variance of the degree distribution can potentially increase final epidemic size. 
In contrast to solutions for unclustered configuration model networks, we find that bond percolation solutions for the final epidemic size are potentially biased if they do not take variable infectious periods into account. 
\end{abstract}
\begin{keyword}
networks \sep SIR \sep clustering \sep percolation
\end{keyword}

\end{frontmatter}

The degree distribution and the clustering coefficient \cite{albert2002statistical} are two of the most commonly investigated features of networks with respect to diffusion and epidemic processes.
We consider the problem of the the dynamic spread of an infectious disease in a population modeled as a random graph.
While the influence of the degree distribution on epidemic dynamics is now a well studied problem\cite{pastor2001epidemic,eames2002modeling,newman2002spread,bansal2007individual}, there is still substantial debate about the effects of clustering. 
Some investigations have indicated that clustering may decrease epidemic thresholds, in effect making it more likely that an epidemic will occur following an initial introduction \cite{newman2003properties}.
Other studies have found that the relationships between clustering and epidemic thresholds is complex \cite{miller2009spread,gleeson2009clustering,PhysRevE.78.048101}, and depends on how clustering is introduced into the population.
The effects of clustering on the timescale of an epidemic are less ambiguous; clustering will decrease the rate of epidemic propagation.
The approach developed here is particularly well suited to investigate the interaction between clustering and the degree distribution in affecting epidemic timescales.
We find that clustering always slows down an epidemic, but the timescale of the epidemic is much more sensitive the variance of the degree distribution.

Following the approach laid out in \citemn, the networks we use are straightforward generalizations of the configuration model \cite{molloy1995critical}.
These networks allow for easy tuning of the degree distribution and the number of 3 cliques in the network, which is related to the clustering coefficient. 
As shown in \citemn, and more recently in \cite{karrer2010random} and \cite{ball2009analysis}, these networks are not locally tree-like, but can nevertheless be analyzed with the tools of branching processes and percolation theory.

Our epidemic model is a generalization of the approach presented in \cite{volz2008sir}, and consists of a low dimensional system of ordinary differential equations which describes the prevalence of infection over time. 
Recently an alternative system of ODEs was independantly developed in \cite{hkmodel} which also describes epidemics in networks with arbitrary degree distribution and clustering.
Our approach complements the one taken in \cite{hkmodel} by providing a solution based on a simple class of random networks that are easily examined using branching processes and percolation theory.
Under some circumstances, our model is in close agreement with the one presented \cite{hkmodel}, but around epidemic thresholds, they can differ substantially. 
This comparison suggests that an accurate account of the effects of clustering requires consideration of more than just the clustering coefficient.
Epidemic dynamics are also affected by details of how clustering is introduced into the network. 

We also revisit the problem of bond percolation for calculating the giant component size in networks with clustering \citemn .
Recently a variation on this class of random networks was investigated in \cite{ball2009analysis}, and they showed how to correctly take the infectious period into account when calculating final epidemic size using bond percolation. 
In unclustered networks generated by the configuration model, bond percolation approximations for final epidemic size that neglect variable infectious periods have negligible bias \cite{kenah2007second,durrett2007random}.
Yet we find that in networks with a lot of clustering, bond percolation approximations for epidemic size can be very biased if variable infectious periods are not taken into account.

\section{Methods}
We consider a basic susceptible-infected-recovered model. 
Infectious nodes transmit to neighbors at a constant rate $\beta$. 
Infectious nodes transition to the recovered state at a constant rate $\gamma$. 
Once recovered, the node cannot be re-infected, and can no longer infect neighbors.

Our solutions will be based on the class of undirected random graphs originally described in \citemn, 
which are refinements of bipartite configuration models \cite{guillaume2003realistic,newman2003properties}. 
A node can be a member of a multiple 2 or 3-cliques ( a \emph{clique} is a completely connected subgraph ). 
A 2 clique is just a pair of nodes with an edge between them, and we will call these \emph{lines}.
A 3 clique is three nodes with all three possible edges, which we will sometimes call \emph{triangles}.
Each node is a member of a random  number of lines and triangles. 
The probability that a node is a member of $l$ lines and $t$ triangles is described by the probability mass function $p_{l,t}$. 
Finally, our model will make heavy use of the probability generating function (pgf) for this distribution:
\[
	g(x, y) = \sum_{l,t} p_{l,t} x^l y^t.
\]

When differentiating the pgf, we will use superscripts so that for example $g^{(x)}$ would indicate the first derivative with respect to $x$ and $g^{(x,x)}$ would indicate the 2nd derivative with respect to $x$. 
The pgf can be used to calculate many useful properties of the graph; for example, the number of half-lines is proportional to $M:= \sum_{l,t} p_{l,t} \times l  = g^{(x)}(1,1)$. 
The number of lines in the network is proportional to $g^{(x)}(1,1)/2$ (because there are two nodes for every line).
And the number of triangles in the network is proportional to $\hat{M}:= g^{(y)}(1,1)/3$ (because there are three nodes for every triangle).

Random graphs of the kind described in \citemn~ can be easily generated by assigning a random number of lines and triangles to a set of $N$ nodes from the distribution $p_{l,t}$. 
After insuring that the mean number of lines and triangles are divisible by 2 and 3 respectively, edges can then be created by
\begin{enumerate}
	\item generating a set of half-lines or ``stubs'', such that the number of times a node appears in the set is equal to the number of lines to which it belongs,
	\item generating a set of ``corners'', such that the number of times a node appears in the set is equal to the number of triangles to which it belongs,
	\item recursively constructing an edge between 2 stubs drawn at random and without replacement,
	\item and, recursively constructing edges between three corners drawn at random and without replacement. 
\end{enumerate}
This algorithm may produce loops and double-edges, but the frequency of such edges will be $O(1/N)$. 

Our solution is based on the idea that the number of transmissions per unit time is a linear function of several time dependent variables:
\begin{enumerate}
	\item $M_{SI}(t) \propto$ the number of lines that begin at a susceptible node and terminate at an infectious node,
	\item $n_{21}(t) \propto$ the number of triangles with two susceptible nodes and one infectious node, 
	\item $n_{12}(t) \propto$ the number of triangles with one susceptible and two infectious nodes, and
	\item $n_{11}(t) \propto$ the number of triangles with one susceptible and one infectious node.
\end{enumerate}
In particular, we will assume that the number of transmissions per unit time over a line and triangle are respectively proportional to
\begin{align}
T_2 &= \beta M_{SI}, \nonumber \\
T_3 &=  \beta (2 n_{21} + 2 n_{12} + n_{11} ). \nonumber
\end{align}
Following the methods in ~\cite{volz2008sir,volz2007susceptible}, we will describe epidemic dynamics with a set of ODEs in terms of the $M$ and $n$ variables as well as two survivor functions: 
\begin{enumerate}
	\item $\theta_2(t)$: the probability that a neighbor in a 2 clique (or ``line'') has not transmitted infection prior to time $t$, and
	\item $\theta_3(t)$: the probability that both neighbors in a 3 clique (or ``triangle'') have not transmitted infection prior to time $t$.
\end{enumerate}
As in \cite{volz2008sir,volz2007susceptible}, we conclude that the probability that a node with $l'$ lines and $t'$ triangles remains susceptible is $\theta_2^{l'} \theta_3^{t'}$.
Consequently the fraction of the population, $S$, that remains susceptible at any time is 
\[
S = \sum_{l,t} p_{l,t} \theta_2^l \theta_3^t = g(\theta_2, \theta_3).
\]

The probability that an edge beginning at a susceptible node will terminate at an infectious node is $M_{SI}/M_S$, where $M_S$ is proportional to the number of half-edges or \emph{stubs} connected to susceptible nodes. 
Similarly, the probability that a susceptible node is connected to a 3 clique with $i$ susceptible nodes and $j$ infectious nodes is $i \times n_{i,j} / \hat{M}_S$, where $\hat{M}_S$ is proportional to the number of links from a susceptible node to a 3 clique. 
These two variables are easily expressed in terms of the pgf:
\begin{align}
M_S = \sum_{l,t} l \times p_{l,t}  \theta_2^l \theta_3^t  = \theta_2 g^{(x)}(\theta_2, \theta_3),   \nonumber \\
\hat{M}_S = \sum_{l,t} t \times p_{l,t} \theta_2^l \theta_3^t = \theta_3 g^{(y)}(\theta_2, \theta_3). \nonumber 
\end{align}
Note that these quantities are less than the actual number by a factor of $1/N$, where $N$ is the population size.

The system of ODEs  relies on several more variables derived from the generating function.
When a transmission event occurs, lines and triangles that were formally counted among $M_{SS}$ or $n_{21}$ may instead be counted among $M_{SI}$ or $n_{12}$. 
Quantifying the magnitude of these changes requires that we calculate the average degree of a newly infected node. 
This is accomplished with the \emph{excess degree distribution} and its corresponding generating function \cite{meyers2006predicting}.
We will denote as $q_{l,t}$ the probability that there are $l$ lines and $t$ triangles connected to a susceptible node that we reach by following a line from an infectious to a susceptible node not counting the line by which we arrived. 
Similarly, $r_{l,t}$ will be the probability that if we follow a 3 clique to a susceptible node, that there are $l$ lines and $t$ 3 cliques  connected to that node, not counting the one by which we arrived. 
Then we have the generating functions 
\begin{align}
g_q(x, y) = \sum_{l,t} q_{l,t} x^l y^t = g^{(x)}(\theta_2 x,\theta_3 y) / g^{(x)}(\theta_2,\theta_3 ) \\
g_r(x, y) = \sum_{l,t} q_{l,t} x^l y^t = g^{(y)}(\theta_2 x,\theta_3 y) / g^{(y)}(\theta_2,\theta_3 ). \nonumber
\end{align}
The mean number of lines and triangles in these joint distributions gives us the expected number of lines or triangles of a newly infected node. 
We denote the means as $\delta_{ij}$, which is the average excess number of type-j links for a susceptible node selected with probability proportional to the number of type-i links. 
Using the generating functions, we have
\begin{align}
\delta_{ll} = \theta_2 g_q^{(x)}(1, 1), \nonumber \\
\delta_{lt} = \theta_3 g_q^{(y)}(1,1), \nonumber \\
\delta_{tt} = \theta_3 g_r^{(y)}(1, 1), \\
\delta_{tl} = \theta_2 g_r^{(y)}(1, 1). \nonumber
\end{align}

The hazard of infection along a single edge is proportional to the probability that the edge terminates at an infectious node ($M_{SI}/M_S$) and the transmission rate. 
As in the equations derived in ~\cite{volz2008sir}, this implies 
\begin{align}
\dot{\theta_3} = -\theta_3 \frac{T_3}{\hat{M}_S}, \nonumber \\
\dot{\theta_2} = -  \theta_2 \frac{T_2}{M_{S}}. \label{eqn:dtheta} \\
\end{align}

Dynamics of $M_{SI}$ and $M_{SS}$ require careful consideration of how edges are rearranged following a transmission event. 
$\dot{M}_{SS} $ describes the time derivative of the (relative) number of lines between susceptibles.
$T_2$ transmissions occur per unit time along lines, and the newly infected individual is connected to an average of $\delta_{ll}$ lines other than the one by which it was infected. 
With probability $M_{SS}/M_S$, one of these lines terminates at a susceptible node. 
Therefore $M_{SS}$ will decrease at a rate of $2 T_2 \delta_{ll} M_{SS}/M_S$. 
Furthermore, $T_3$ transmissions will occur via triangles, and the newly infected node will be connected to an expected number $\delta_{tl}$ lines.
Each of these will also terminate at a susceptible node with probability $M_{SS}/M_S$. 
Then we conclude
\begin{align}
\label{eqn:dmss}
\dot{M}_{SS} = - 2 \frac{M_{SS}}{M_S} \left( T_2 \delta_{ll}   +  T_3 \delta_{tl}  \right).
\end{align}

Similar reasoning leads to the equation for $\dot{M}_{SI}$. 
The edge rearrangement follows a similar patterns as for $M_{SS}$, but we must also account for the \emph{increase} of $M_{SI}$ when a newly infected node is connected to another susceptible (with probability $M_{SS}/M_S$), as well as the \emph{decrease} when the new infection has connections to other infecteds (with probability $M_{SI}/M_S$). 
Taking this into account yields terms of the form $ (T_2 \delta_{ll} + T_3 \delta_{tl} ) M_{XY}/M_X $. 
And, in addition to the edge-rearrangement terms, we must account for changes due to recovery ($-\gamma M_{SI}$) and direct transmission ($-\beta M_{SI}$). 
\begin{align}
\label{eqn:dmsi}
\dot{M}_{SI} = -M_{SI} (\gamma + \beta)  + \left(T_2 \delta_{ll} + T_3 \delta_{tl} \right) \left( \frac{M_{SS}}{M_S} - \frac{M_{SI}}{M_S}  \right)
\end{align}

Finally, the equations for the number of 3 cliques with $i$ susceptible and $j$ infectious constituents, $n_{ij}$, is found by considering rearrangements as above, as well as flow between classes that are due to an infectious member of the 3 clique transmitting to a susceptible member, or recovering. 
For example, a 3 clique with one susceptible and two infectious nodes (state $(1,2)$) will transition to the state $(0,3)$ at the rate $2 \beta$, since there are two edges between susceptible and infecteds in this clique. 
And it will transition to the state $(1,1)$ at the rate $2 \gamma$, since there are two infectious nodes in the clique that can recover. 
To summarize, we find

\begin{align}
\dot{n}_{30} &= - \left(T_3 \delta_{tt} + T_2 \delta_{lt} \right) \frac{3 n_{30}}{\hat{M}_S},  \nonumber \\
\dot{n}_{21} &= - (2 \beta  + \gamma) n_{21} +  \left(T_3 \delta_{tt} + T_2 \delta_{lt} \right) \left(\frac{3 n_{30}}{\hat{M}_S} -\frac{2 n_{21}}{\hat{M}_S} \right),  \nonumber \\
\dot{n}_{20} &= \gamma n_{21}  -\left(T_3 \delta_{tt} + T_2 \delta_{lt} \right) \frac{2 n_{20}}{\hat{M}_S},   \label{eqn:dn}\\  
\dot{n}_{12} &= 2 \beta n_{21} - (2 \beta  + 2\gamma) n_{12} +  \left(T_3 \delta_{tt} + T_2 \delta_{lt} \right) \left(\frac{2 n_{21}}{\hat{M}_S} -\frac{n_{12}}{\hat{M}_S} \right),  \nonumber \\ 
\dot{n}_{11} &=  2 \gamma n_{12} - ( \beta   + \gamma) n_{11}  +  \left(T_3 \delta_{tt} + T_2 \delta_{lt} \right) \left(\frac{2 n_{20}}{\hat{M}_S} -\frac{n_{11}}{\hat{M}_S} \right).  \nonumber 
\end{align}

An additional differential equation can be solved for the epidemic prevalence at any time.
\begin{align}
\dot{I} &= -\dot{S} - \gamma I \nonumber \\
 &= \frac{d}{dt} g(\theta_2, \theta_3) - \gamma I \nonumber \\
  &= \dot{\theta}_2 g^{(x)}(\theta_2, \theta_3) + \dot{\theta}_3 g^{(y)}(\theta_2, \theta_3) - \gamma I 
\end{align}

If an initial fraction $\epsilon \ll 1$ of the population is infected at the beginning of the epidemic, we use the initial conditions
\begin{align}
\theta_3(0) &= \epsilon   \nonumber \\
\theta_2(0) &= \epsilon \nonumber \\
M_{SI}(0) &= \epsilon M   \\
M_{SS}(0) &= (1-2 \epsilon) M  \nonumber \\
n_{30}(0) &= (1 - \epsilon) \hat{M}  \nonumber \\
n_{21}(0) &= \epsilon \hat{M}, \nonumber
\end{align}
and the remaining variables would be zero. 

\subsection{Bond percolation}
	Bond percolation solutions for the giant component size were independently derived in \citemn. 
	These solutions were based on the undirected bond percolation, such that each edge is ``occupied'', or transmits infection, with indepandent probability $\tau$. 
	Setting the edge occupation probability to the transmission probability per partnership, which yields $\tau = \beta/(\beta+\gamma)$ in the Poisson process considered here, the giant component size approximately corresponds to the final size of an epidemic under certain conditions \cite{meyers2006predicting} (the correspondence is exact when the infectious period is constant).
	Derivations were also provided in \citemn~ for the probability that an epidemic will occur following a single introduction in clustered networks, and the threshold transmissibility $\tau$ for epidemics to be possible (i.e. where a giant component forms). 
	But these derivations did not account for variable infectious periods in a realistic epidemiological setting. 
	It is now understood that the undirected bond percolation giant component size is only an approximation for the final epidemic size when the infectious period is not constant, albeit a very good approximation ~\cite{durrett2007random, kenah2007second}. 
	And the bond percolation solutions for the probability of an epidemic can be very biased. 
	
	In contrast to what is observed in \cite{kenah2007second}, we find that the bond percolation solution for final size is not always a good approximation in networks with clustering.
	Recentlly an alternative percolation technique was developed in \cite{ball2009analysis}  which correctly accounts for variable infectious periods and can accurately calculate final sizes in clustered networks, and the techniques described in \cite{miller2009spread} could also take this into account. 
	
	The solutions in \citemn~ were based on the calculation of the probability that there would be 0, 1 or 2 secondary infections following an initial infection in a 3 clique. 
	If the infectious period is $t$ and assuming a constant rate of transmission, the transmission probability to one neighbor would be $\tau(t) = 1- e^{-\beta t}$. 
	We will denote the mean transmissibility as $\bar{\tau} = \beta / (\beta  + \gamma)$. 
	The bond percolation solutions in \citemn~ were based on the idea that each edge is occupied with independent probability $\bar{\tau}$, which implies that the probability of having 1 or 2 secondary infections in a 3 clique is 
	\begin{itemize}
	\item 1 secondary infection: $\bar{\alpha}_1 := 2 \bar{\tau} (1-\bar{\tau})^2$, \\
	\item 2 secondary infections: $\bar{\alpha}_2 := \bar{\tau}^2 + 2 \bar{\tau}^2 (1-\bar{\tau})  $.
	\end{itemize}	
	In fact, these probabilities are functions of the infectious period of the initial case in the 3 clique, which is itself an exponentially distributed random variable. 
	We can solve for the true probabilities by integrating over the infectious period. 
	\begin{itemize}
	\item 1 secondary infection: 
	\begin{align}
		\alpha_1 &:= \int_0^\infty \gamma e^{-\gamma t} \left( 2 (1-e^{-\beta t}) e^{-\beta t} (1-\bar{\tau}) \right)  \mathrm{d}t \nonumber \\
		& = 2  (1-\bar{\tau} )^2 - 2 \frac{\gamma}{2 \beta + \gamma} (1-\bar{\tau} ) \nonumber
	\end{align}
	\item 2 secondary infections:
	\begin{align}
		\alpha_2 & := \int_0^\infty \gamma e^{-\gamma t}  \left(  (1-e^{-\beta t})^2 + 2 (1-e^{-\beta t}) e^{-\beta t} \bar{\tau}  \right) \mathrm{d}t \nonumber \\
		 & = 1 + (1-2 \bar{\tau}) \frac{\gamma}{2\beta + \gamma} + 2(1-\bar{\tau})(\bar{\tau}-1)  \nonumber
	\end{align}
	\end{itemize}
	This distribution is generally different from the one based on  $\bar{\alpha}_1$ and $\bar{\alpha}_2$, and the expected number of secondary infections is strictly less with variable infectious periods. 
	To see this, denote the averages $R = 2 \alpha_2 + \alpha_1$ and $\bar{R} = 2 \bar{\alpha_2} + \bar{\alpha}_1$, 
	and note that only 2nd order terms of $\tau$ will differ between $R$ and $\bar{R}$. 
	We have $\bar{\tau}^2 = \beta^2 / (\beta + \gamma)^2$, and 
	\begin{align}
	\langle \tau^2 \rangle &= \int_0^\infty \gamma e^{-\gamma t} ( 1-e^{-\beta t})^2 \mathrm{d} t \nonumber \\
	 &= \frac{2\beta^2}{ (\beta + \gamma)(2\beta + \gamma) }
	\end{align}
	It is easy to see that $\langle \tau^2 \rangle > \bar{t}^2$.
	Furthermore, if we collect all terms involving $\tau^2$ in the equation for $R$, we find a leading factor of $-2\bar{\tau}$. 
	Consequently, these terms will be negative and will have larger magnitude in the expression for $R$ than for $\bar{R}$, so $R < \bar{R}$. 
	
	Below, we consider an approximate bond percolation calculation for final size. 
	We propose that the number of secondary infections in each 3 clique to which an infected belongs is generate by $1 - \alpha_1 - \alpha_2 + \alpha_1 x + \alpha_2 x^2$, with the probabilities $\alpha$ calculated above that take the infectious period into account.
	With this modification, the giant component size can be calculated as in \cite{2009arXiv0904.3253M}, however it is still only an approximation because it assumes that the number of secondary infections is independent in multiple cliques to which an infected belongs. 
	In general, because $R<\bar{R}$, this solution will underestimate final size, while the solution in \citemn~ will overestimate final size.

\subsection{Alternative models}
We compare solutions of the system ~\ref{eqn:dtheta}-\ref{eqn:dn} to stochastic simulations in continuous time.
The simulations are based on the the Gillespie algorithm \cite{gillespie1977exact}. 
Random  networks are generated as described above. 
At time $t=0$, A number $I(0)$ initial infections are selected uniformly at random in the graph.
When a susceptible is infected, new transmission and recovery events are queued with exponentially distributed waiting times. 

We also compare the clustering model to a recently proposed system of ODEs based on moment-closure \cite{hkmodel}. 
This model was developed for networks with a given degree distribution generated by $G(x)$ and a clustering coefficient $\phi$. 
This system does not specify a joint distribution for the number of lines and triangles. 
Rather, this system is based on the idea that potential 3 cliques, of which a degree $k$ node will have ${k \choose 2}$, will exist with independent probability $\phi$. 
This system also uses PGFs within a low-dimensional system of ODEs,  and proposes that $S = G(\theta)$, with $\dot{\theta} = - \theta \beta [SI]/M_S$, where $[SI]$ is the number of half-edges from a susceptible node that terminates at an infectious node. 
Equations for $[SI]$ are derived in terms of the number of connected triples, or 2-paths, of nodes which pass through a susceptible. 
This model makes the approximation that the number of 2-paths connecting two susceptibles and an infected is a simple function of the clustering coefficient $\phi$: 
\[
	[SSI] \approx [SS][SI] \frac{G''(\theta) }{N (G'(\theta))^2}  \left( (1-\phi) + \phi G'(1) \frac{[SI]}{\theta G'(\theta) M_I} \right) .
\]
 And the number of 2-paths connecting a susceptible with two infecteds is 
\[
	[ISI] \approx [SI]^2 \frac{G''(\theta) }{N (G'(\theta))^2}  \left( (1-\phi) + \phi G'(1) N \frac{[II]}{M^2_I} \right) .
\]
We will subsequently refer to this as the House-Keeling (HK) model. 

\section{Results}

Many of our results were generated for the purpose of explicating the interaction between the variance of the degree distribution and the level of clustering.
The clustering model is especially well suited for investigating this problem, since it remains low dimensional even with many degree classes.
Previous research has elucidated the importance of the variance of the degree distribution on the location of epidemic thresholds and the final size ~\cite{pastor2001epidemic, newman2002spread}. 
In particular, this research has shown that in networks with power law degree distributions, as the variance of the degree distribution diverges to infinity, so too does the reproduction number, and the threshold transmissibility vanishes. 

To simultaneously investigate heterogeneous degree distributions and clustering, we constructed a degree distribution based on the negative binomial (NB) distribution which allows us to  hold the mean of the distribution constant while interpolating over a wide range of variance
bound between the mean of the distribution and infinity.
The NB distribution with parameters $p$ and $r$ is generated by
\begin{align}
g_{nb}(x ; r,p) = \left( \frac{p}{1 - (1-p) x} \right)^r. 
\label{eqn:gnb0}
\end{align}
We will modify this distribution so that an expected fraction $p_t$ of links are to 3-cliques and 2-cliques always appear in pairs. 
Each edge will occur as part of a pair, which may form part of a 3 clique (with probability $p_t$), or may simply be a pair of edges with two nodes that are not themselves connected. 
Then given a random number $k$ 2-tuples generated by equation \ref{eqn:gnb0}, the number of lines and triangles is generated by $((1-p_t) x^2 + p_t y)^k$, where $y$ is the dummy variable for 3 cliques, and $x$ is the dummy variable for 2 cliques. 
Using the composition property of pgf's, the degree distribution is generated by
\begin{align}
g(x,y) = g_{nb}((1-p_t) x^2 + p_t y). \label{eqn:gnb}
\end{align}
Since lines always appear in pairs, it is easy to keep the mean of the distribution constant while tuning the amount of clustering with $p_t$, which can range between zero and one. 

\begin{figure}
\includegraphics[width=\textwidth]{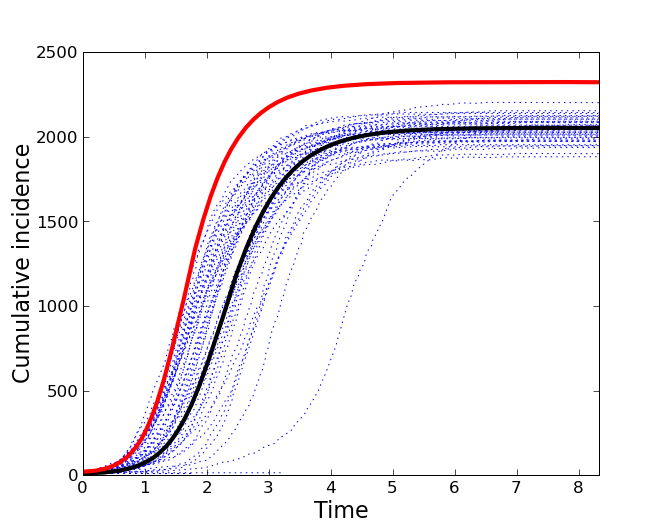}
\caption{The cumulative number of infections is shown versus time.
Fifty stochastic simulations are compared to the solution of equations \ref{eqn:dtheta}-\ref{eqn:dn}. 
The degree distribution is generated by equation \ref{eqn:gnb}. 
$N = 5000, I(0)=10, p_t = 0.9, \beta = 1,\textrm{and } \gamma = 1 $.
For comparison, a trajectory with $p_t = 0$ is shown in red.
\label{fig:cumPrev}}
\end{figure}

Figure ~\ref{fig:cumPrev} shows a comparison of the clustering model to 50 stochastic simulations on random networks with 5000 nodes and 10 initial infections. 
The degree distribution was generated by equation~\ref{eqn:gnb}, with a mean of 2 and a variance of 3. 
The fraction of links to 3 cliques was $p_t = 90\%$. 
For comparison, we also plot a solution to the clustering model (red line) with $p_T = 0$, so that there is no clustering. 
Clustering has the effect of slowing down the epidemic and reducing the final number ultimately infected. 
And the system of equations \ref{eqn:dtheta}--\ref{eqn:dn} correctly predicts the final size, while the trajectory passes through the central mass of simulated trajectories; the analytical model approximately corresponds to the median time for a stochastic simulation to reach a given prevalence (results not shown).

\begin{figure}
\includegraphics[width=\textwidth]{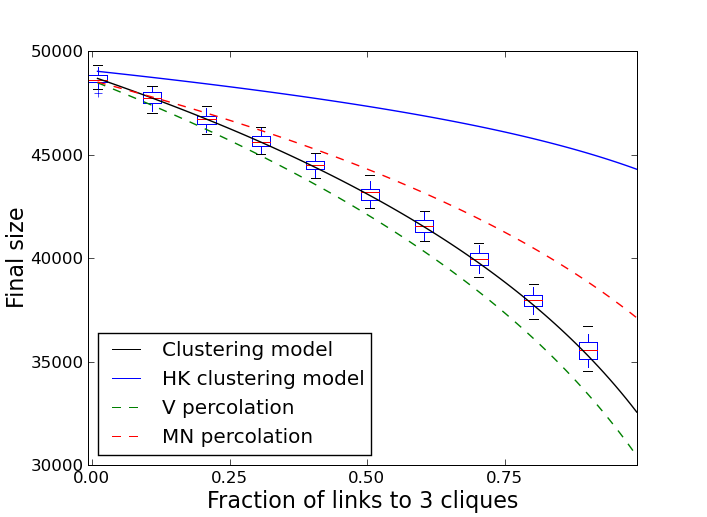}
\caption{A comparison of the final epidemic size versus the level of clustering in the network. 
The degree distribution is generated by equation~\ref{eqn:gnb}. 
The black line corresponds to the solution of equations \ref{eqn:dtheta}--\ref{eqn:dn}. 
The boxplots illustrate the 90\% confidence interval from 50 stochastic simulations on networks with 5000 nodes. 
$\beta = \gamma = 1$, and both the mean and variance of the degree distribution is 2. 
\label{fig:fsSim}}
\end{figure}

We show the effects of clustering on the final size of the epidemic in figure \ref{fig:fsSim}. 
The clustering model (black line, equations \ref{eqn:dtheta}--\ref{eqn:dn}) correctly reproduce the final epidemic size observed in simulations (box plots).
The MN percolation solution proposed in \citemn (red dashed line) is noticeably biased for non-zero clustering, although that solution does trend downwards correctly. 
Over-estimation by the MN model is expected, as in the last section we showed that the MN model will overestimate the number of secondary infections within a 3 clique when the infectious period is not constant. 

To calibrate the HK model with our chosen $p_t$, we used the univariate generating function
\[
G(x) = g(x, x^2), 
\]
since there are two edges for every 3-clique.
And we set $\phi$ to be the clustering coefficient in the network, which as shown in \citemn~ is the ratio of $3\times$ the number of triangles, which we denote $N_{Delta}$, to the number of 2-paths in the network, which we denote $N_3$.
We have
\begin{align}
\phi &= 3 N_\Delta / N_3 \nonumber \\
  &= \frac{g^{(y)}(1, 1)}{\frac{1}{2} G''(1)}.
\end{align}
The HK clustering model also overestimates final size; this is not unexpected, since the HK model is not premised on the introduction of 3 cliques, but is rather an approximation in the event that the every potential triangle exists with independent probability $\phi$. 
The lack of alignment of the HK model and equations~\ref{eqn:dtheta}--\ref{eqn:dn} indicates that accurate accounting of the effects of clustering requires consideration not only of macroscopic properties like the clustering coefficient, but also the details of how clustering is introduced into the network. 
Furthermore, as we show below, the discrepancy between the HK and clustering model is greatest when the variance of the degree distribution is low, which is the case in Figure \ref{fig:fsSim}; the variance is set at its lower bound, and is equal to the mean=2. 

\begin{figure}
\includegraphics[width=1.2\textwidth]{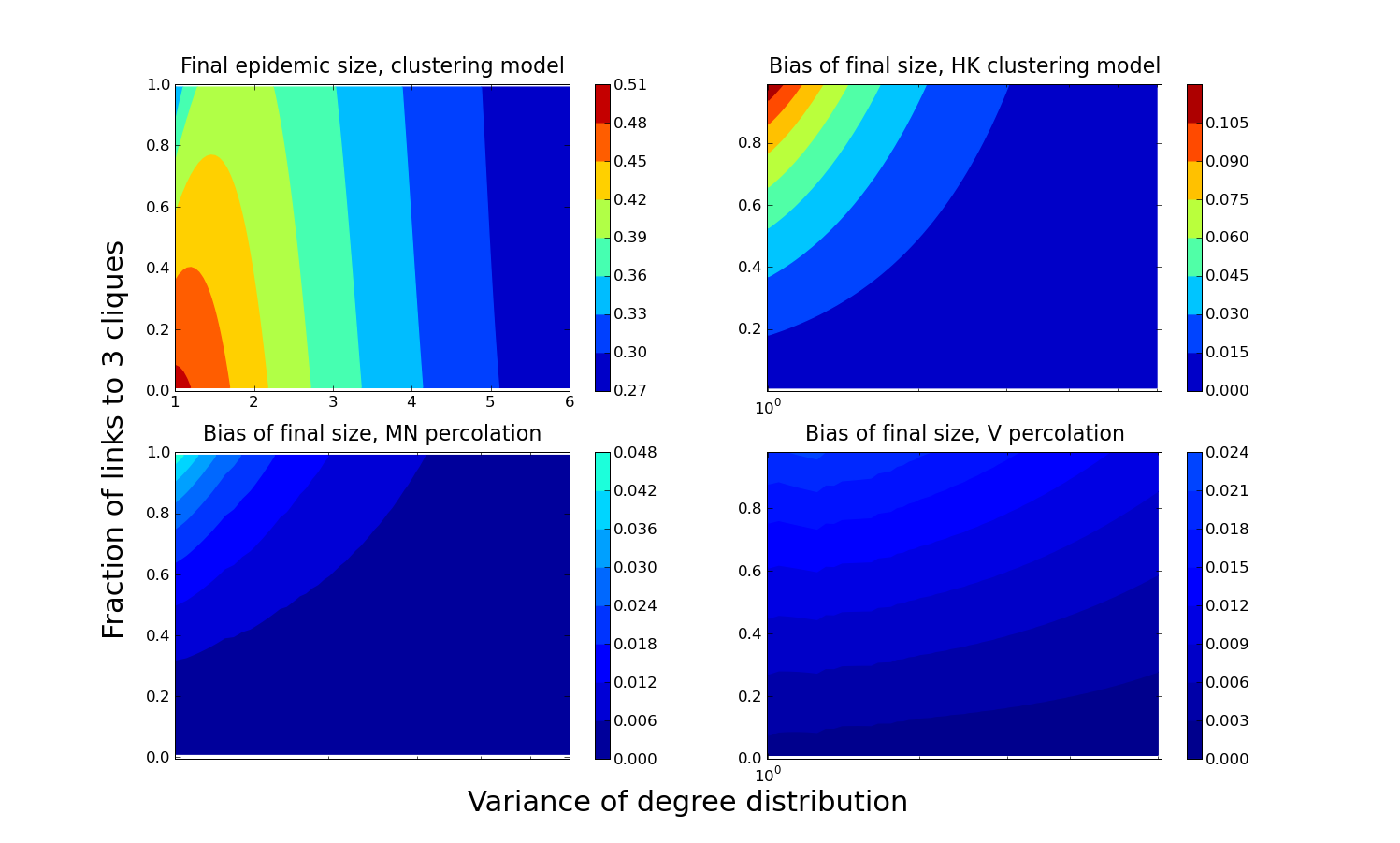}
\caption{Upper left: The final epidemic size is shown as a function of clustering (the fraction of links that go to 3 cliques) and the variance of the degree distribution. 
All results are based on a solution of equations \cm . $\beta = \gamma = 1$. 
The remaining panels show the difference between the clustering model (\cm) and an alternative solution for the final size. \label{fig:fsAnalyt}}
\end{figure}

A more detailed characterization of the interplay of variance and clustering is illustrated in Figure~\ref{fig:fsAnalyt}. 
The upper left panel shows the final size predicted by the clustering model (equations \ref{eqn:dtheta}--\ref{eqn:dn}). 
As expected, final size decreases with increasing clustering (the fraction of links to 3 cliques).
And in agreement with previous studies, the final size usually decreases with increasing variance. 
There is an exception, however, when the variance is very small, and clustering is high. 
In this region, with variance between 1 and 1.5, we see that final size can actually increase with larger variance. 

The remaining panels in Figure~\ref{fig:fsAnalyt} show the discrepancy between the clustering model and an alternative calculation of final size. 
These heatmaps were calibrated to have the same color scale. 
The bias increases with clustering in all cases. 
But bias is insubstantial when the variance is large, even if clustering is also large.
This can partially be explained by noting that the nonlinear relationship between $p_t$ and the clustering coefficient.
Given a constant fraction of 3 cliques, $p_t$, the number of 3 cliques in the network is
\[
N_\Delta = \sum_{l,t} t\times p_{l,t}/3 =  g^{(y)}(1, 1)/3, 
\]
which is clearly constant with respect to the variance of the degree distribution (holding the mean constant).
But the number of 2 paths is
\[
N_3 = \frac{1}{2} G''(1) = \sum_{k = l+2t} p_{k = l + 2t} {k \choose 2},
\]
which clearly increase with the second moment of the distribution ($\sum_k p_k k^2 $). 
So, increasing the variance of the distribution (holding the mean constant) has the effect of decreasing the ratio of $N_\Delta$ to $N_3$. 
The clustering coefficient, $\phi = 3 N_\Delta / N_3$ is more important than the total number of 3 cliques in determining epidemic outcomes, and as we increase variance, $\phi$ converges to zero, and the clustering model converges to the percolation and HK model solutions. 
To understand why $\phi$  rather than $N_\Delta$ is the important quantity for determining final size, note that as we increase the variance of the degree distribution, the mean excess degree, $G''(1) / G'(1)$, increases. 
The number of 2 paths through a node of degree $k$ is ${k \choose 2}$. 
So if we consider a node with mean excess degree $k = G''(1)/G'(1)$, which is the mean degree of a new infected early in the epidemic, the probability that two neighbors of that node are themselves connected is
\[
\frac{p_t}{ k-1} = \frac{p_t G'(1)}{G''(1)}, 
\]
which will decrease with variance of the degree distribution. 

\begin{figure}
\includegraphics[width=1.2\textwidth]{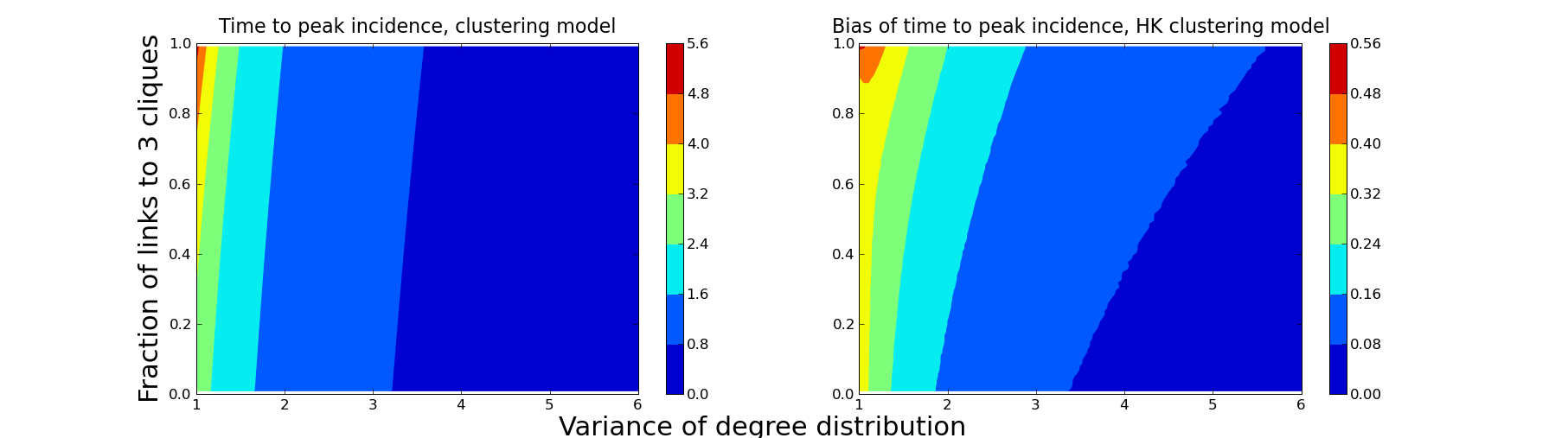}
\caption{Left: The time to peak epidemic incidence ($t_p$) as a function of the fraction of links to 3 cliques, $p_t$, and the variance of the degree distribution. 
Results are based on a solution to equations \cm, with $\beta = \gamma = 1$ and a degree distribution generate by equation ~\ref{eqn:gnb}.
Right: The discrepancy between the $t_p$ predicted by the HK model and the clustering model (equations \cm). \label{fig:ttpAnalyt}}
\end{figure}

Figure ~\ref{fig:ttpAnalyt} illustrates the impact of clustering and variance of the degree distribution on dynamical aspects of the epidemic. 
We consider the time to peak incidence, defined as $t_p = \mathrm{argmax}(-\dot{S}(t))$ . 
Clustering always has the effect of slowing down the epidemic and increasing $t_p$. 
Variance always has the effect of speeding up an epidemic and decreasing $t_p$. 
Of the two, it appears that $t_p$ is much more elastic with respect to variance than $p_t$. 
The HK model is in close agreement with the clustering model (equations \cm), but can differ by as much as $10\%$ when $p_t$ is large. 
The percolation methods are not represented in this figure, as they are uninformative about the timescale of the epidemic.

\section{Conclusion}
The model presented here is a generalization of the one presented in ~\cite{volz2008sir}. 
Several other generalizations of that model have been presented \cite{2009arXiv0909.4485M}, including 
simultaneous network dynamics, such as edge swapping \cite{volz2007susceptible,volz2009epidemic}, 
populations with heterogeneous contact rates\cite{volz2008susceptible},
multiple edge types with distinct transmission rates \cite{volzEpidemics}, 
preferential attachment \cite{volzEpidemics}, 
and growing networks with natural birth and mortality \cite{kamp2009untangling}. 
It is likely that these approaches could be combined with the model presented here. 
For example, it would be straightforwards to consider epidemics in clustered networks that also have dynamically rearranging ties.
Other straightforward generalizations might include larger clique sizes, or the inclusion of network motifs other than cliques. 
This model should be generalizable to the more general class of bipartite random network presented in \cite{karrer2010random}. 

Regarding the potential for empirical applications, 
network samples increasingly provide the information necessary to parameterize these models.
Degree distributions and clustering coefficients are often ascertained in social network studies \cite{rothenberg2000atlanta,abramovitz2009using}.
Epidemiological surveillance data often also provide partnership durations and measures of concurrency \cite{volzEpidemics,foxman2006measures}.

The problem of SIR dynamics in clustered networks with arbitrary degree distributions was also investigated in a recent manuscript ~\cite{hkmodel} (the HK model). 
We have compared that model to ours by calibrating the clustering coefficient $\phi$ to match the fraction of links to 3 cliques, $p_t$. 
The models are in close agreement when the variance of the degree distribution is high, but substantial differences in both the final size and timescale of the epidemic exist when the degree distribution is homogeneous and when $p_t$ is large. 
This comparison shows that epidemic dynamics depend on more than just the clustering coefficient, but on details of how clustering is introduced into the network. 
While the HK model is more parsimonious than the one presented here (it has fewer variables), we speculate that our model will be a good alternative for data with well defined cliques, such as human populations with household structure \cite{longini1982household,ball2009analysis}.

\emph{Acknowledgements:} The author acknowledges support from NIH U01 GM087719. 
The author thanks Thomas House and Joel Miller for valuable feedback. 

\bibliographystyle{elsarticle-num} %
\bibliography{clusteringReferences}

\end{document}